# JWST/MIRI Observations of PAH Emission and Evolution in H II Regions of NGC 5457

Atul Kumar Singh, Rahul Kumar Anand, Shantanu Rastogi

*Department of Physics, Deen Dayal Upadhyaya Gorakhpur University,
Civil Lines, Gorakhpur-273009*

## ABSTRACT

**Aims:** To investigate the physical processing of PAHs in high-mass star-forming environments. This study aims to characterize how the PAH ionization fraction varies across different H II regions and to search for spectroscopic evidence of the processing or destruction of PAH molecules in intense radiation fields.

**Methodology:** We utilized mid-infrared integral field spectroscopic data from the JWST's Mid-Infrared Instrument (MIRI) in its Medium Resolution Spectroscopy (MRS) mode. One-dimensional spectra were extracted for each of the four H II regions. We measured the integrated fluxes of the prominent PAH emission features at 7.7, 8.6, and 11.3 µm and calculated key diagnostic flux ratios to probe the physical state of the PAH population.

**Results:** The four H II regions in NGC 5457 exhibit significant diversity in their PAH spectral characteristics. The diagnostic $F(8.6)/F(11.3)$ ratio, a tracer of PAH ionization, varies from 0.23 to 0.62 across the sample. The region with the highest ionization (highest $F(8.6)/F(11.3)$) shows the lowest $F(7.7)/F(8.6)$ ratio (1.13), while the region with the lowest ionization has the highest $F(7.7)/F(8.6)$ ratio (2.38).

**Conclusion:** The observed spectral diversity is primarily driven by variations in the PAH ionization state, which is governed by the local radiation field intensity. The strong anti-correlation between the $F(7.7)/F(8.6)$ and $F(8.6)/F(11.3)$ ratios provides a new, powerful diagnostic for PAH processing. This trend is interpreted as an evolutionary sequence where increasingly harsh radiation fields not only ionize the PAH population but also alter the surviving cations, consistent with the selective destruction of the carriers of the 7.7 µm feature.

\* E-mail address: shantanu_r@hotmail.com

# 1. INTRODUCTION

Polycyclic aromatic hydrocarbons are a ubiquitous and fundamental component of the interstellar medium in star-forming galaxies (Khramtsova et al., 2013; Li, 2020; Murata et al., 2017; Ujjwal et al., 2024). These large, carbonaceous molecules, thought to consist of tens to hundreds of carbon atoms (Murata et al., 2017; Pathak & Rastogi, 2006; Pathak & Rastogi, 2008), are the primary carriers of the prominent mid-infrared emission features observed at 3.3, 6.2, 7.7, 8.6, 11.3, 12.7, and 17 μm (Tielens, 2008; Li, 2020; Maragkoudakis et al., 2023; Rigopoulou et al., 2024). Collectively, the luminosity from these bands can account for up to 20% of a galaxy's total infrared output (Hensley et al., 2022; Sandstrom et al., 2010). The emission mechanism is understood to be a process of stochastic heating (Draine & Li, 2001; Pavlyuchenkov et al., 2012): a single ultraviolet or optical photon from nearby stars is absorbed by a PAH molecule, which is rapidly heated to a high internal temperature. The molecule then relaxes through a cascade of vibrational transitions, emitting photons at characteristic MIR wavelengths corresponding to its fundamental C-C and C-H stretching and bending modes (Draine et al., 2021; Sandstrom et al., 2010).

The diagnostic power of these features is immense. The relative strengths and detailed profiles of the PAH bands serve as sensitive probes of the local physical conditions within the ISM (Knight et al., 2021; Maragkoudakis et al., 2022; Peeters et al., 2002). For instance, the features at 3.3 and 11.3 μm are attributed to C-H out-of-plane bending modes in neutral PAHs (Mackie et al., 2015; Maltseva et al., 2016), while the 6.2 and 7.7 μm features are primarily associated with C-C stretching modes in ionized PAHs (cations) (Dontot et al., 2020; Wenzel et al., 2025). Consequently, ratios such as the 11.3/7.7 μm flux ratio can effectively trace the PAH ionization fraction, providing a window into the charge state of the interstellar grain population (Maragkoudakis et al., 2022; Sidhu et al., 2022). Similarly, the relative strength of the 3.3 μm feature, which arises from smaller PAHs, compared to the longer-wavelength features, which are more prominent in larger molecules, can be used to diagnose the average size of the PAH population (Maragkoudakis et al., 2020; Maragkoudakis, Peeters, & Ricca, 2023; Anand et al., 2025). Despite this rich diagnostic potential, our understanding of the PAH life cycle has been historically limited by observational constraints. Spatially resolving these variations within diverse galactic environments, from the harsh, ionized interiors of HII regions to the more quiescent photodissociation regions that envelop them, has been challenging, preventing the construction of a comprehensive, observationally grounded model of how PAH properties are shaped by their surroundings (Knight et al., 2021; Zang et al., 2022).

The advent of the James Webb Space Telescope has inaugurated a new era in MIR astronomy, providing the necessary tools to overcome these longstanding limitations (Labiano et al., 2021; Wright et al., 2023). Specifically, the Mid-Infrared Instrument and its Medium Resolution Spectrometer offer an unprecedented combination of sensitivity, sub-arcsecond spatial resolution (0.2"-0.7"), and integral field spectroscopic capability (Dicken et al., 2024; Labiano et al., 2021; Patapis et al., 2023). This represents a monumental leap beyond previous facilities like the Spitzer Space Telescope, as MIRI MRS offers significantly improved spatial resolution and sensitivity (Bouwman et al., 2023; Morrison et al., 2023). The MRS enables the dissection of individual star-forming clumps and the detailed mapping of physical conditions on scales of tens of parsecs within nearby galaxies (Rigopoulou et al., 2024; Ujjwal et al., 2024). A critical design feature of the MRS is its continuous wavelength coverage, spanning 4.9 to 27.9 μm at a resolving power of R ~ 3700-1300 (Argyriou et al., 2023). This is sufficient to resolve the intrinsic profiles of the broad PAH features and deconvolve them from underlying dust continua and nearby fine-structure lines. Most importantly, it encompasses the entire set of major PAH emission features in a single observation, allowing for a robust and systematic study of their relative variations without the systematic uncertainties inherent in stitching together data from different instruments or observing modes. This provides a powerful and self-consistent lever for untangling their complex excitation mechanisms and environmental dependencies (García-Bernete et al., 2022; Rigopoulou et al., 2024).

To leverage these transformative capabilities, we have targeted NGC 5457 (M101, the Pinwheel Galaxy), a nearby (~6.7 Mpc), grand-design spiral galaxy of type SAB(rs)cd. Its nearly face-on orientation (~18°) minimizes projection effects and geometrical ambiguities, which would otherwise complicate the interpretation of line-of-sight emission, making it an ideal laboratory for spatially resolved spectroscopy. The galaxy hosts a well-cataloged population of HII regions (C. & Garnett, 1996), presenting a rich and statistically significant sample of star formation activity across a full range of galactic environments, from the inner disk to the outer arms (Hu et al., 2018; Linden et al., 2022; Pleuss et al., 2000). Furthermore, a wealth of extensive ancillary data is available for NGC 5457, including direct-method metallicity measurements from the CHAOS project, which have established a distinct radial abundance gradient across its disk (Esteban et al., 2019; Galliano, 2011; Kang et al., 2021; Whitcomb et al., 2024). This gradient is particularly valuable, as both the formation of PAHs (requiring carbon) and their destruction (via hard UV photons in low-metallicity environments) are intimately linked to metallicity (Gordon et al., 2008; Khramtsova et al., 2013; Maragkoudakis et al., 2023). The established gradient thus provides a natural experiment, allowing for the study of how PAH properties correlate with the chemical evolution of a galaxy, all within a single, self-contained system (Lin, Zou, Kong, Lin, Mao, Cheng, et al., 2013; Lin, Zou, Kong, Lin, Mao, Jiang, et al., 2013).

In this paper, we present a detailed analysis of JWST/MIRI MRS observations targeting four distinct HII regions within NGC 5457. These regions were selected to sample a diverse range of local environments, from compact, high-excitation star formation knots likely powered by very young, massive star clusters to more evolved, extended complexes where stellar feedback has had more time to impact the surrounding medium. Our primary scientific goal is to conduct a rigorous test of theoretical models of PAH processing. We will use key diagnostic ratios, including those involving the 7.7, 8.6 and 11.3 μm features, to create spatially resolved maps of the PAH size and ionization fraction (Maragkoudakis et al., 2022). We seek to answer specific questions, such as: How does the PAH ionization fraction vary with proximity to the central ionizing source (Goicoechea et al., 2009)? Do we observe evidence for the destruction of small PAHs in the harshest radiation fields (Egorov et al., 2023; Galliano,2011; Khramtsova et al., 2014; Micelotta et al., 2009, 2010; Monfredini et al., 2019; Murata et al., 2017; Pavlyuchenkov et al., 2013; Siebenmorgen, 2009, Anand et al., 2023)?

## 2. OBSERVATIONS AND DATA REDUCTION

This study is based on archival data from the James Webb Space Telescope (JWST) obtained from the Mikulski Archive for Space Telescopes (MAST). The analysis focuses on mid-infrared integral field spectroscopy of four HII regions within the nearby spiral galaxy NGC 5457 (M101) (Rogers et al., 2023).

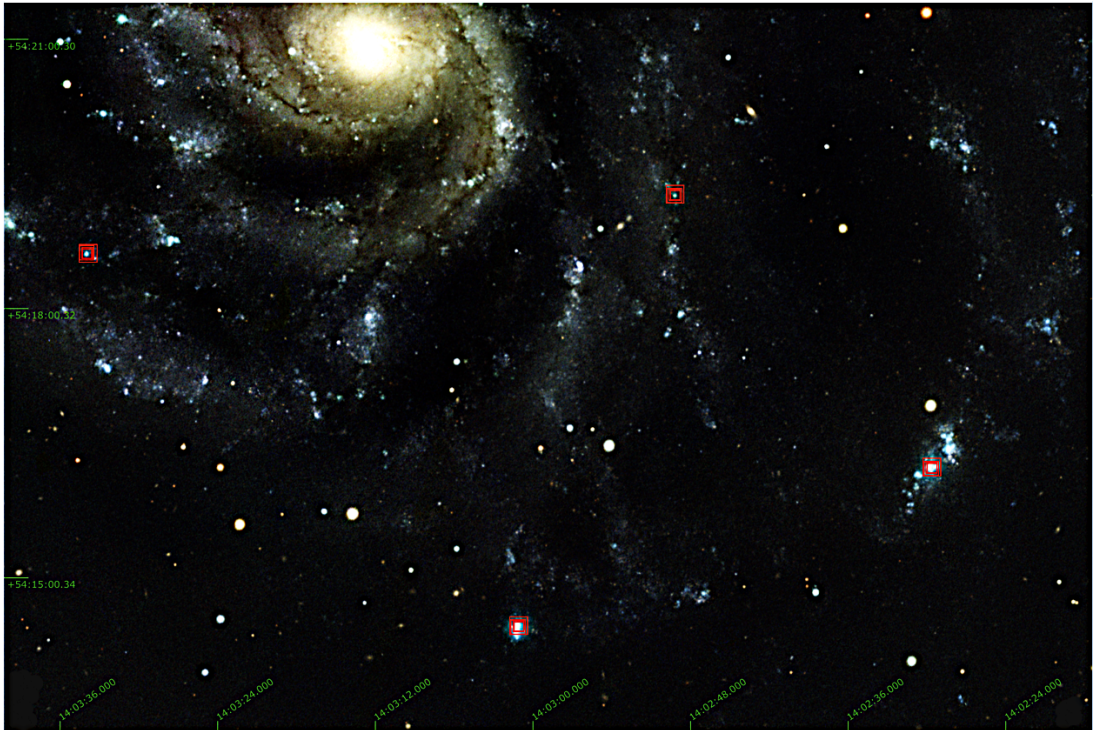

Figure 1: The locations of our four program targets in the grand-design spiral galaxy NGC 5457 (M101). The red squares indicate the positions of the HII regions: NGC5457-205-98, NGC5457+189-136, NGC5457-100-388, and NGC5457-361-280. The background is a public-release color composite image from the Pan-STARRS1 (PS1) survey. The axes are labeled with J2000 Right Ascension and Declination. In this orientation, north is up and east is to the left.

## 2.1. JWST MIRI Observations

The targets of this investigation are four distinct HII regions within the grand-design spiral galaxy NGC 5457, a prototypical Sc I galaxy located at a distance of approximately 7 Mpc. The specific regions, designated NGC5457-205-98, NGC5457+189-136, NGC5457-100-388, and NGC5457-361-280, were selected to sample a range of physical conditions. A summary of the observations, including the target coordinates and the associated JWST General Observer (GO) program identifiers, is provided in Table 1.

Table 1: Coordinates and JWST proposal identification for the NGC 5457 regions analyzed in this study. Right ascension (R.A.) and declination (Dec.) are given in J2000 coordinates.

| Target Designation | R.A. | Dec. | JWST Proposal ID |
|---|---|---|---|
| NGC5457-205-98 | 14:03:05.8 | +54:20:56.0 | 4297 |
| NGC5457+189-136 | 14:03:13.5 | +54:22:12.0 | 4297 |
| NGC5457-100-388 | 14:03:20.2 | +54:18:08.0 | 4297 |
| NGC5457-361-280 | 14:03:28.9 | +54:19:16.0 | 4297 |

The observations were performed using the Mid-Infrared Instrument in its Medium Resolution Spectroscopy mode (Wright et al., 2023). The MRS is a powerful integral field spectrograph that provides simultaneous spatial and spectral information over the 4.9 to 27.9 µm wavelength range, a spectral window inaccessible to other JWST instruments at this resolving power (Argyriou et al., 2023). The instrument utilizes four separate Integral Field Units, designated as Channels 1, 2, 3, and 4 (Argyriou et al., 2023), which are fed by a series of dichroics to observe four distinct wavelength intervals simultaneously. Each IFU employs an image slicer to reformat a two-dimensional field of view into a long slit, which is then dispersed by a grating onto a 1024x1024 Si:As detector array (Morrison et al., 2023).

The archival data has wavelength ranges of interest for this study; the observations were executed using the SHORT (A) and LONG (C) grating settings for MIRI Channels 1 and 2. This configuration results in 4 distinct spectral bands: 1A, 1C, 2A, 2C (Argyriou et al., 2023). The spectral resolving power, $R=\lambda/\Delta\lambda$, is wavelength-dependent, varying from $R\approx3,500$ at the shortest wavelengths in Channel 1 to $R\approx2,400$ at the wavelengths in Channel 3 (Argyriou et al., 2023). The field of view also increases with wavelength, from approximately 3.2″×3.7″ in Channel 1 to 4.0″×4.8″ in Channel 2 (Patapis et al., 2023). A summary of the instrumental parameters for each spectral band used is presented in Table 2.

Table 2: Spectral coverage, field of view, and resolving power for the instrument channels and sub-bands used in this work.

| Channel | Sub-band | Wavelength Range (µm) | Nominal FOV (arcsec) | Mean Resolving Power (R = $\lambda/\Delta\lambda$) |
|---|---|---|---|---|
| 1 | LONG (C) | 6.53–7.65 | 3.2 × 3.7 | 3,355 |
| 2 | SHORT (A) | 7.51–8.77 | 4.0 × 4.8 | 3,050 |
| 2 | LONG (C) | 10.01–11.70 | 4.0 × 4.8 | 3,080 |

A standard 4-point dither pattern was employed for all observations. This observing strategy is critical for several reasons. First, it allows for the mitigation of detector artifacts, such as bad pixels and residual cosmic ray hits, by observing the target at multiple detector locations. Second, and more fundamentally, it is required to properly sample the instrument's point spread function (PSF) and line spread function (LSF). Both the spatial and spectral resolution elements of JWST are intentionally undersampled by the MIRI detector pixels; the use of a dither pattern with sub-pixel offsets allows for the reconstruction of a well-sampled, high-fidelity data cube during the final stages of the data reduction pipeline. The archival FITS headers indicate that Target Acquisition (TA) was performed for each pointing, which refines the pointing accuracy from the nominal JWST blind pointing uncertainty of ~0.3" to a precision of approximately 90 mas, ensuring the compact HII region cores were accurately centered within the IFU field of view.

### 2.2. Data Retrieval and Pipeline Processing

The data were retrieved programmatically from the MAST portal using the astroquery Python package. The analysis is based on the Level 3 science products, which are generated by the official JWST Science Calibration Pipeline. The specific data products used were processed with pipeline *version 1.17.1 (Build 11.2)* and the Calibration Reference Data System (CRDS) context *jwst_1413.pmap*. Specifying the pipeline version and CRDS context is of fundamental importance for the reproducibility of this work. The calibration of MIRI, particularly at longer wavelengths, is known to be time-dependent due to a gradual decrease in detector throughput, which is most pronounced in Channels 3 and 4. The JWST pipeline incorporates a time-dependent sensitivity model to correct for this effect, and this model is updated with new pipeline builds. Using data processed with a specified, recent build ensures that the flux calibration is as accurate as possible and that the results can be reliably compared to other studies.

The JWST pipeline processing is divided into three principal stages:

***Stage 1 (calwebb_detector1):*** This stage operates on the raw telemetry data (*uncal.fits), which consists of up-the-ramp integrations. It performs fundamental detector-level corrections, including subtraction of a super bias frame, flagging of saturated pixels, linearity correction, and dark current subtraction. A key step is the identification and flagging of jumps in the ramps caused by cosmic ray impacts. The final output is a set of two-dimensional slope images (*rate.fits) in units of DN/s for each dither position.

***Stage 2 (calwebb_spec2):*** This stage takes the individual slope images and applies a series of spectroscopic calibrations. For MRS data, this includes applying a flat-field correction to account for pixel-to-pixel sensitivity variations, correcting for stray light, and applying a first-pass fringe correction. Crucially, this stage also performs the photometric (flux) calibration, converting the signal from DN/s to physical units (MJy/sr), and attaches a World Coordinate System (WCS) solution to the data. The WCS provides the mapping from detector pixel coordinates to on-sky position (RA, Dec) and wavelength for every pixel in the IFU. The wavelength calibration is based on in-flight observations of giant planets and atmospheric models, achieving a typical accuracy of a few km/s. The output products are fully calibrated individual exposures (*cal.fits).

***Stage 3 (calwebb_spec3):*** This final stage combines the multiple dithered exposures from Stage 2 into a single, science-ready data product. The process begins with background matching and outlier detection, which identifies and rejects residual bad pixels and cosmic rays that were not flagged in Stage 1. The core of this stage is the resampling of the data from the individual, distorted detector frames onto a regular, three-dimensional grid with two spatial axes (RA, Dec) and one spectral axis (wavelength). This step, often referred to as "cube building," uses a 3D drizzle-like algorithm to combine the dithered pointings, resulting in a final data cube (*s3d.fits) for each of the 12 spectral bands.

### 2.3. Spectral Extraction and Final Data Products

The quantitative analysis in this paper is based on the Level 3, three-dimensional data cubes (*s3d.fits) produced by the *calwebb_spec3* pipeline. The final spectral extraction and analysis were performed using the Jdaviz software package, specifically its Cubeviz configuration.

For each of the six spectral bands per target, the corresponding *s3d.fits data cube was loaded into the Cubeviz application. Visual inspection of the collapsed "white-light" image in the Cubeviz viewer confirmed that the HII regions are compact and well-centered. To extract a one-dimensional spectrum, we utilized the interactive tools within Cubeviz. A circular spatial region was defined in the image viewer to encompass the bright core of each HII region. The one-dimensional spectrum was then generated using the "3D Spectral Extraction" plugin. We selected the 'Sum' function to perform aperture photometry within the defined circular region at each wavelength slice of the cube. The plugin propagates the per-voxel uncertainties from the array of the input data cube to generate a final 1D error spectrum. The resulting 1D spectrum for each band, in units of surface brightness (MJy/sr), was then used for subsequent analysis within the tool.

### 2.4 Spectroscopic Analysis

The spectroscopic analysis of the extracted 1D spectra was conducted interactively using the analysis plugins available within Cubeviz, which leverages the specutils package. For the broad Polycyclic Aromatic Hydrocarbon (PAH) features, the underlying continuum was modeled and subtracted prior to flux measurement. This was achieved by fitting a Spline3 to anchor points selected in adjacent, feature-free spectral regions.

## 3. RESULTS

This section presents the primary observational findings derived from the JWST/MIRI MRS data cubes for the four targeted H II regions in NGC 5457. The focus is on the quantitative characterization of the mid-infrared spectra, particularly the prominent PAH emission features.

### 3.1. Mid-Infrared Spectra of the Target H II Regions

The analysis is based on the integrated one-dimensional spectra extracted from the core of each of the four H II regions: NGC5457-100-388, NGC5457-205-98, NGC5457-361-280, and NGC5457+189-136. The resulting spectra, covering the wavelength range from approximately 6.5 to 11.7 µm, are presented in Figures 2, 3, 4, and 5.

A visual inspection of these figures reveals the characteristic signatures of PAH emission in all four targets. The spectra are dominated by strong, broad emission features, which stand out prominently against the underlying dust continuum. Consistent with extensive previous observations of star-forming regions across the universe, we clearly detect the well-known PAH features centered near 7.7, 8.6, and 11.3 µm. These features are understood to arise from the radiative relaxation of stochastically heated PAH molecules, with the 7.7 µm complex primarily attributed to C-C stretching modes, the 8.6 µm feature to C-H in-plane bending modes, and the 11.3 µm feature to C-H out-of-plane bending modes. The strong presence of these bands confirms that a significant population of these large carbonaceous molecules permeates the interstellar medium (ISM) within these active star-forming sites.

While all four regions exhibit these characteristic features, there are notable qualitative differences in their spectral morphology. The relative intensity of the 11.3 µm feature compared to the 7.7 µm complex, for instance, appears to vary substantially among the targets. In NGC5457-361-280 (Figure 4), the 11.3 µm feature is particularly strong, appearing as the dominant feature in this spectral window. In contrast, for NGC5457+189-136 (Figure 5), the 7.7 µm complex and the 11.3 µm feature have more comparable peak fluxes. These visual variations strongly suggest that the underlying physical conditions and the properties of the PAH populations are not uniform across our sample, a conclusion that will be quantified in the following section.

### 3.2. PAH Feature Fluxes and Diagnostic Ratios

To move beyond a qualitative description, the integrated fluxes of the 7.7, 8.6, and 11.3 µm PAH features were measured for each target after subtracting a local continuum. These measurements allow for the calculation of diagnostic flux ratios, which are powerful probes of the physical state of the PAH population. The measured fluxes and a set of key diagnostic ratios are compiled in Table 3.

The data presented in Table 3 reveal significant, quantitative variations in the properties of the PAH emission across the four H II regions. The flux ratios, which normalize out differences in the total PAH abundance and overall brightness of the regions, exhibit a considerable range.

The F(7.7)/F(8.6) ratio varies from a low of 1.13 in NGC5457+189-136 to a high of 2.38 in NGC5457-361-280, a factor of more than two. Even more striking is the variation in the F(8.6)/F(11.3) ratio, which spans a range from 0.23 to 0.62, a factor of nearly three.

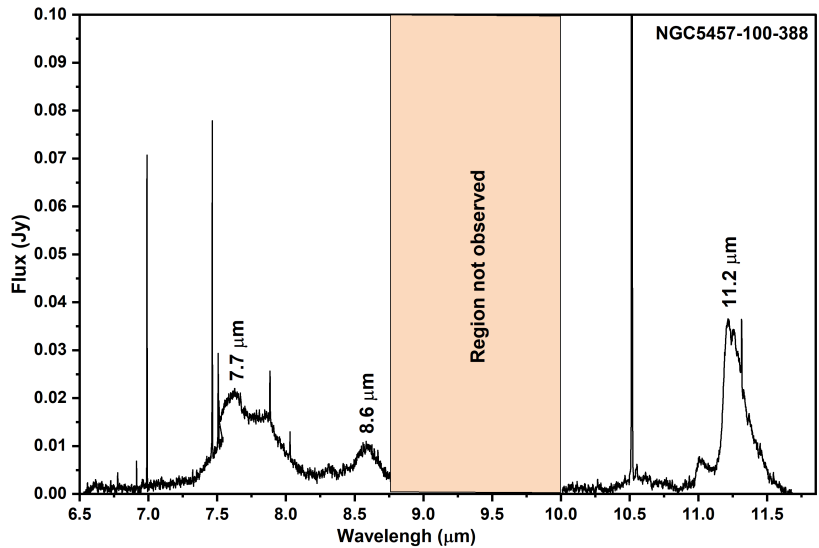

Figure 2: JWST/MIRI spectrum of NGC5457-100-388 from 6.5 to 11.7 µm. Key Polycyclic Aromatic Hydrocarbon (PAH) emission features are identified. The shaded region marks a gap in the observational data.

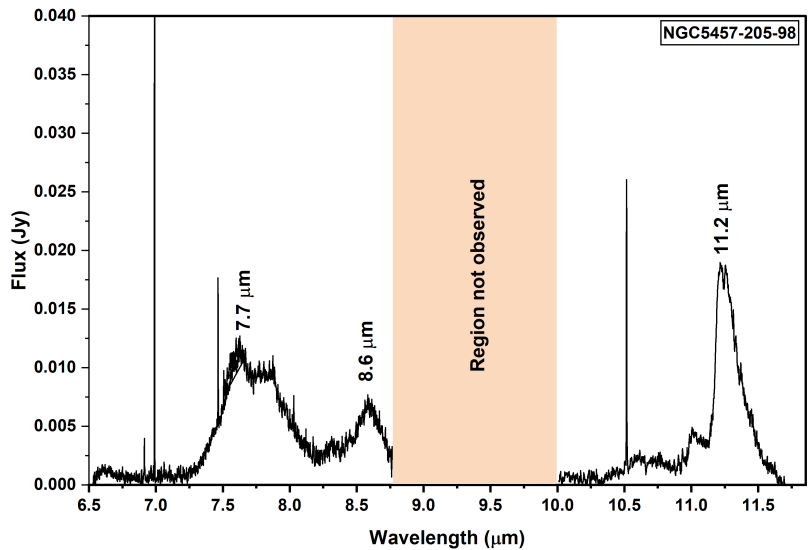

Figure 3: JWST/MIRI spectrum of NGC5457-205-98 from 6.5 to 11.7 µm. Key Polycyclic Aromatic Hydrocarbon (PAH) emission features are identified. The shaded region marks a gap in the observational data.

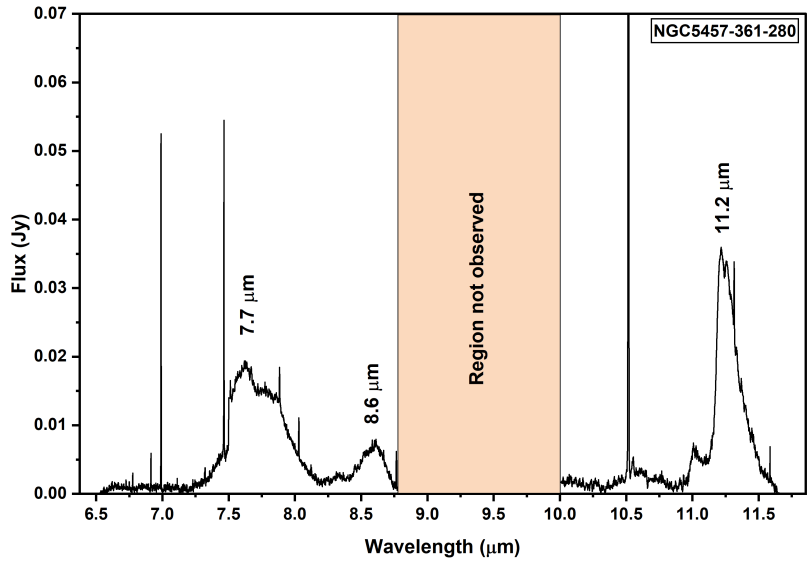

*Figure 4: JWST/MIRI spectrum of NGC5457-361-280 from 6.5 to 11.7 µm. Key Polycyclic Aromatic Hydrocarbon (PAH) emission features are identified. The shaded region marks a gap in the observational data.*

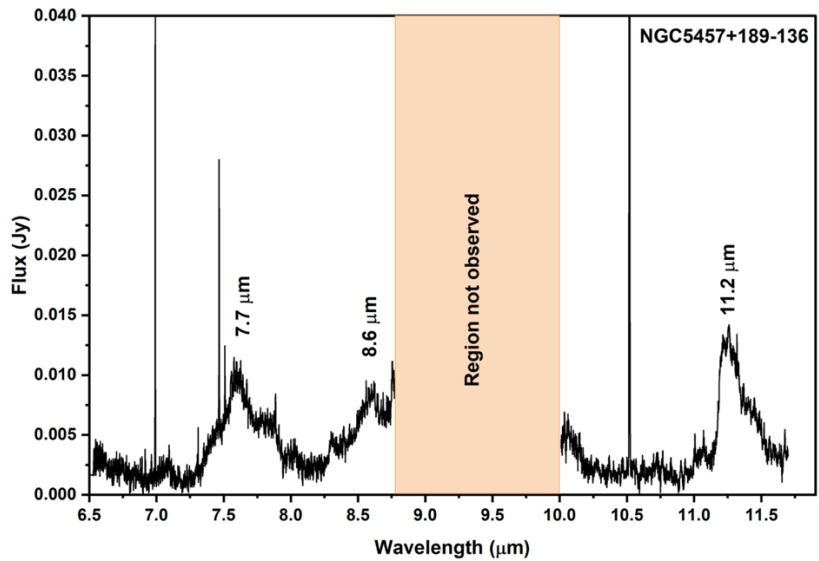

*Figure 5: JWST/MIRI spectrum of NGC5457+189-136 from 6.5 to 11.7 µm. Key Polycyclic Aromatic Hydrocarbon (PAH) emission features are identified. The shaded region marks a gap in the observational data.*

A key finding from this quantitative analysis is the existence of a strong anti-correlation between the F(7.7)/F(8.6) and F(8.6)/F(11.3) ratios within our sample. The H II region exhibiting the highest F(8.6)/F(11.3) ratio (NGC5457+189-136) simultaneously shows the lowest F(7.7)/F(8.6) ratio. Conversely, the region with the lowest F(8.6)/F(11.3) ratio (NGC5457-361-280) displays the highest F(7.7)/F(8.6) ratio. The other two regions, NGC5457-100-388 and NGC5457-205-98, lie at intermediate positions in this trend. This systematic relationship suggests that a single, dominant physical parameter or process is responsible for driving the observed variations in the PAH spectra across these diverse star-forming environments. The physical interpretation of this trend forms the core of the subsequent discussion.

Table 3: Measured PAH fluxes at 7.7, 8.6, and 11.3 µm and their corresponding band ratios for selected regions in NGC 5457. Fluxes are in arbitrary units consistent with Figures 2–5. The ratios provide a diagnostic of PAH ionization state, with higher F(7.7)/F(11.3) values generally indicating a larger fraction of ionized PAHs.

| Object | F(7.7 µm) | F(8.6 µm) | F(11.3 µm) | F(7.7)/F(11.3) | F(8.6)/F(11.3) | F(7.7)/F(8.6) |
|---|---|---|---|---|---|---|
| NGC5457-100-388 | 0.020 | 0.009 | 0.036 | 0.56 | 0.25 | 2.22 |
| NGC5457-205-98 | 0.011 | 0.007 | 0.019 | 0.58 | 0.37 | 1.57 |
| NGC5457-361-280 | 0.019 | 0.008 | 0.035 | 0.54 | 0.23 | 2.38 |
| NGC5457+189-136 | 0.009 | 0.008 | 0.013 | 0.69 | 0.62 | 1.13 |

## 4. DISCUSSION

The quantitative results presented in the previous section reveal a clear and systematic variation in the PAH spectral characteristics across the four targeted H II regions. In this section, these results are interpreted within the established theoretical framework of PAH emission to diagnose the physical conditions within these regions. This analysis aims to construct a coherent physical picture that addresses the primary scientific questions of this study regarding the processing of PAHs in high-mass star-forming environments.

### 4.1. The Physical Interpretation of PAH Band Ratios: Probing the PAH Ionization State

The diagnostic power of PAH features stems from the fact that the relative strengths of different vibrational modes are highly sensitive to the intrinsic properties of the molecules, primarily their charge state and size, and the nature of the exciting radiation field.
A broad consensus, built upon decades of laboratory experiments, quantum-chemical calculations, and astronomical observations, holds that the charge state of the PAH population is the principal driver of the large-scale variations seen in mid-infrared spectra (Bauschlicher et al., 2008; Rigopoulou et al., 2024). Specifically, the emission features in the 6–9 µm range, which include the prominent 7.7 µm and 8.6 µm bands, are known to be significantly enhanced in ionized PAHs (cations) relative to their neutral counterparts (Maragkoudakis et al., 2022; Sakon et al., 2011; Sidhu et al., 2022, Maurya et al., 2023). These features arise from C-C stretching and C-H in-plane bending modes whose oscillator strengths increase dramatically upon ionization (Dontot et al., 2020; Esposito et al., 2024; Mackie et al., 2015; Semmler et al., 1991).

Conversely, features arising from C-H out-of-plane bending modes, most notably the strong band at 11.3 µm, are the dominant spectral signature of neutral PAHs (Candian & Sarre, 2015; Mackie et al., 2015; Maltseva et al., 2016; Sakon et al., 2011). The intensity of this feature is much stronger relative to the 6–9 µm complex in neutral species (Maragkoudakis et al., 2022; Sidhu et al., 2022).

This dichotomy provides a powerful diagnostic tool. Ratios of a "cationic" band flux to a "neutral" band flux, such as F(7.7)/F(11.3) or F(8.6)/F(11.3), serve as effective tracers of the average ionization fraction of the PAH population (Maragkoudakis et al., 2022; Sakon et al.,

2011; Sidhu et al., 2022). A higher ratio implies a larger fraction of ionized PAHs, which in turn points to an environment with a higher ionization parameter, defined by the ratio of the incident far-ultraviolet photon flux to the local gas density (Maragkoudakis et al., 2022; Sidhu et al., 2022).

Applying this diagnostic to the data in Table 3 allows for the establishment of an ionization sequence among the four target regions. NGC5457+189-136 exhibits the highest values for both F(7.7)/F(11.3) (0.69) and F(8.6)/F(11.3) (0.62), indicating that it harbors the most highly ionized PAH population in our sample. This is characteristic of an environment exposed to a particularly intense and/or hard radiation field. At the other extreme, NGC5457-361-280 shows the lowest values for these ratios (0.54 and 0.23, respectively), signifying a predominantly neutral PAH population, typical of a more shielded or quiescent environment like a photodissociation region (Goicoechea et al., 2009; Knight et al., 2021; Sidhu et al., 2022, Maurya & Rastogi, 2015). The remaining two H II regions, NGC5457-100-388 and NGC5457-205-98, occupy intermediate positions in this ionization sequence. This variation confirms that our sample successfully probes a diverse range of radiation environments within NGC 5457.

### 4.2. A Diagnostic Diagram of PAH Environments in NGC 5457

To visualize the relationships between the PAH properties and to better interpret the observed trends, we construct a diagnostic diagram using the measured flux ratios (Galliano et al., 2008; Maragkoudakis et al., 2023). Such diagrams are powerful tools for separating astronomical sources based on the physical state of their ISM (Galliano et al., 2008).

We propose a diagram plotting the F(7.7)/F(8.6) ratio against the F(8.6)/F(11.3) ratio, as shown conceptually in Figure 6. The x-axis, F(8.6)/F(11.3), serves as our primary proxy for the PAH ionization state, with values increasing to the right for more highly ionized environments subject to more intense FUV radiation fields (Galliano et al., 2008; Maragkoudakis et al., 2022). The y-axis, F(7.7)/F(8.6), probes the intrinsic properties of the PAH cation population (Knight et al., 2021).

When the data for the four H II regions from Table 3 are plotted on this diagram, they do not scatter randomly. Instead, they form a remarkably clear sequence extending from the upper-left to the lower-right of the plot space. NGC5457-361-280, with the lowest ionization and highest F(7.7)/F(8.6) ratio, anchors the upper-left of this sequence. NGC5457+189-136, with the highest ionization and lowest F(7.7)/F(8.6) ratio, lies at the lower-right terminus. The other two sources, NGC5457-205-98 and NGC5457-100-388, fall neatly along the track connecting these two extremes.

This diagram effectively segregates the four H II regions according to the physical state of their PAH populations (Galliano et al., 2008; Maragkoudakis et al., 2022). The tight correlation between the two ratios provides strong evidence that a single physical gradient is responsible for the full range of observed spectral properties (Maragkoudakis et al., 2023). This gradient, which simultaneously increases PAH ionization while altering the emission characteristics of the cations themselves, can be interpreted as a sequence of increasing environmental harshness, driven by the radiation fields of the embedded massive stars (Khramtsova et al., 2014; Knight, Peeters, Tielens, et al., 2021; Knight, Peeters, Wolfire, et al., 2021).

### 4.3. An Evolutionary Sequence of PAH Processing in H II Regions

The clear trend observed in the diagnostic diagram (Figure 6) allows us to construct a physical narrative of PAH processing. The sequence of H II regions across the diagram is not random but likely represents an environmental or evolutionary sequence, tracing the transformation of

PAHs as they are exposed to increasingly intense and hard radiation fields. This framework provides direct, observationally grounded answers to the core scientific questions of our study.

The physical mechanism proceeds as follows. The ionization state of the PAH population, traced by our x-axis F(8.6)/F(11.3), is set by the local ionization parameter, which increases with proximity to young, massive star clusters that produce copious FUV photons. The observed anti-correlation implies that the same process driving ionization—exposure to energetic photons—also systematically lowers the F(7.7)/F(8.6) ratio.

We can interpret the two endpoints of our observed sequence as follows:

NGC5457-361-280 and NGC5457-100-388 (Upper-Left; Low Ionization): These regions represent the relatively "unprocessed" or shielded end of the sequence. Their low F(8.6)/F(11.3) ratios (0.23 and 0.25) are indicative of a predominantly neutral PAH population. Such conditions are characteristic of PDRs that surround H II regions, or more evolved, lower-density H II regions where the radiation field has been diluted or softened. In this environment, the F(7.7)/F(8.6) ratio is high (2.38 and 2.22). This can be considered the "native" ratio for a population of PAH cations excited by a relatively benign FUV field.

NGC5457+189-136 (Lower-Right; High Ionization): This region represents the "highly processed" end of the sequence. Its very high F(8.6)/F(11.3) ratio (0.62) points to an extreme environment where PAHs are almost completely ionized. This requires direct exposure to the intense, hard FUV radiation field emanating from a very young, compact, massive star cluster at the heart of the H II region. Recent JWST studies have confirmed that such regions with high ionization parameters are sites of significant PAH processing and destruction. The key finding of this work is the concurrent, dramatic decrease in the F(7.7)/F(8.6) ratio to 1.13 in this harsh environment. This provides a new spectral signature for PAH processing. The energetic photons responsible for ionizing the PAHs must also be altering the cation population itself. This could occur via two primary mechanisms:

Selective Destruction: The carriers of the 7.7 μm feature may be systematically smaller or structurally less stable than the carriers of the 8.6 μm feature. In the most intense radiation fields, these more fragile cations are preferentially destroyed, leading to a decrease in the F(7.7)/F(8.6) ratio. This aligns with broader findings that the smallest PAHs are destroyed in the hardest radiation fields.

Excitation Effects: Alternatively, the intrinsic emissivity ratio of the C-C stretching (7.7 μm) and C-H in-plane bending (8.6 μm) modes within a given PAH cation may change as a function of the absorbed photon energy. Excitation by very hard photons could alter the vibrational cascade, channeling less energy through the 7.7 μm mode relative to the 8.6 μm mode.

Regardless of the precise microphysical cause, the observed trend strongly suggests that as the radiation field becomes more intense, it not only ionizes the PAH population but also fundamentally processes the surviving cations. NGC5457-205-98 represents an intermediate case, lying along this processing track and demonstrating the continuous nature of these environmental effects.

**4.4 The Spectral Character of HII Regions in NGC 5457: Archetypal Class A Population**

The mid-infrared spectra observed in this study, particularly from the most highly ionized regions such as NGC5457+189-136, are characterized by a 7.7 μm emission complex that clearly peaks shortward of 7.7 μm, with a prominent sub-feature near 7.6 μm. This spectral morphology provides a direct connection to the empirical classification scheme developed by Peeters et al. (2002) from a large sample of ISO spectra.

In their seminal work, Peeters et al. (2002) classified sources based on the profiles and peak positions of the 6.2, 7.7, and 8.6 µm features.

Class A sources are defined by a 6.2 µm feature peaking near 6.22 µm and, most relevant to this study, a 7.7 µm complex dominated by the 7.6 µm component (designated Class A'). A key conclusion from their survey was that Class A spectra are archetypal for interstellar medium (ISM) environments, such as HII regions and reflection nebulae. The spectra from the HII regions in NGC 5457 are therefore quintessential examples of the Peeters et al. Class A population. This placement is significant for two reasons: first, it validates our observations as being consistent with the known properties of high-mass star-forming environments; second, it establishes a critical link to the physical interpretations associated with this spectral class.

The work of Peeters et al. (2002) established this environmental link by comparing a diverse collection of objects across the sky. The present study provides a high-fidelity, spatially resolved confirmation of this connection. By demonstrating that this classification holds true even when examining the diversity. Within the HII region population of a single galaxy, this work strengthens the conclusion that the physical environment of an HII region naturally gives rise to Class A PAH spectra.

**4.5 The Physical Nature of the Class A Spectrum: Probing PAH Size and Charge**

The link between environment and spectral class can be understood through theoretical models that explore the dependence of PAH emission on molecular properties. The theoretical work of Pathak & Rastogi (2008) provides a compelling physical explanation for the Class A spectra observed in the NGC 5457 HII regions. Using quantum-chemical calculations, they modeled the emission from different populations of PAHs, varying their size and charge state. A key finding was that the observed 7.7 µm complex, with its sub-features at 7.6 µm and 7.8 µm, is a direct diagnostic of the underlying PAH population.

Specifically, their Models I and II, which simulate emission from populations of small-to-medium-sized (≤40 carbon atoms) PAH cations, successfully reproduce a spectrum where the 7.6 µm component is dominant. This profile matches the Class A' spectra defined by Peeters et al. (2002) and observed in our target HII regions. In contrast, their Model III, representing large PAH cations (>40 C atoms), produces a spectrum dominated by the 7.8 µm feature, which they associate with more "benign" environments such as the outflows of carbon-rich stars.

This theoretical framework allows for a unified physical narrative that connects the observations of Peeters et al. (2002), the models of Pathak & Rastogi (2008), and the results of the present study. The intense and hard far-ultraviolet (FUV) radiation fields characteristic of HII regions are responsible for two concurrent effects that shape the local PAH population:

1. Ionization: The FUV photons ionize the PAHs, a necessary condition for activating the strong C-C stretching modes that produce the entire 7.7 µm complex.

2. Size Selection: The same energetic photons can preferentially destroy larger, less stable PAHs through photodissociation, leaving behind a resilient population of more compact, small-to-medium-sized cations.

This process naturally selects for the exact type of PAH population, ionized and medium-sized, that the models of Pathak & Rastogi (2008) predict will produce a 7.6 μm-dominant spectrum. The observations of Peeters et al. (2002) provide the empirical "what" (HII regions have Class A spectra), Pathak & Rastogi (2008) provide the theoretical "why" (Class A spectra arise from medium-sized cations), and the present work provides the direct observational evidence of "how" the local radiation field within HII regions drives the ionization and likely the size selection that produces this signature. This cohesive picture is summarized in Table 4.

Table 4: Comparison of PAH Spectral Properties in NGC 5457 with Observational Classes and Theoretical Models

| Object/Model | Dominant 7.7 μm Component | Ionization Tracer F(8.6)/F(11.3) | Processing Tracer F(7.7)/F(8.6) | Inferred PAH Properties (Size, Charge) | Associated Environment |
|---|---|---|---|---|---|
| NGC5457+189-136 (This Work) | 7.6 μm (Class A-like) | 0.62 (High) | 1.13 (Low) | Small/Medium Cations, Highly Processed | Harsh, High-Ionization HII Region Core |
| NGC5457-361-280 (This Work) | 7.6 μm (Class A-like) | 0.23 (Low) | 2.38 (High) | Small/Medium Cations, Less Processed | Shielded/Evolved HII Region/PDR |
| Peeters et al. (2002) Class A | 7.6 μm | High (implied) | N/A | Substituted/Complexed PAHs | HII Regions, RNe, Galaxies |
| Peeters et al. (2002) Class B | ~7.8 μm | Low (implied) | N/A | Pure, Large PAHs | PNe, Isolated Herbig AeBe Stars |
| Pathak & R. (2008) Model II | 7.6 μm | High (Cations) | N/A | Medium Cations (20-40 C) | UV-rich Environments (HII Regions) |
| Pathak & R. (2008) Model III | 7.8 μm | High (Cations) | N/A | Large Cations (>40 C) | Benign Regions (Stellar Outflows) |

## 4.6. Broader Context and Implications

These findings, enabled by the unprecedented spectroscopic capabilities of JWST/MIRI, provide a detailed view of PAH physics within individual star-forming regions. This work complements and adds depth to large-scale multi-wavelength surveys.

Recent results from the PHANGS-JWST program, which uses broadband filters to map entire galaxies, have similarly found that H II regions host PAH populations that are, on average, more highly ionized than those in the diffuse ISM. Our spectroscopic study drills down into individual H II regions, demonstrating that there is significant diversity.

The host galaxy, NGC 5457, is known to possess a significant radial metallicity gradient. Metallicity is another key parameter expected to influence PAH properties, with low-metallicity environments thought to favor smaller and more neutral PAHs, potentially due to a harder interstellar radiation field. While our sample of four regions is too small to disentangle the effects of the local radiation field from the global metallicity gradient, this remains a critical goal for the field. Future studies with larger samples of H II regions spanning a wide range of galactocentric distances will be essential to isolate these two effects.

This study set out to answer two specific questions. We can now provide clear, observationally motivated answers:

How does the PAH ionization fraction vary? Our results demonstrate that the ionization fraction varies dramatically among H II regions, from predominantly neutral to almost fully ionized. This variation, traced by the F(8.6)/F(11.3) ratio, is the primary driver of the observed spectral diversity and is likely governed by the intensity and hardness of the local FUV radiation field from embedded star clusters.

Do we observe evidence for the destruction of small PAHs? Yes, this work provides a new spectral tracer of PAH processing. The systematic decrease in the F(7.7)/F(8.6) ratio in lockstep with increasing ionization provides compelling evidence that the PAH cation population is being fundamentally altered in the harshest radiation fields, consistent with the selective processing or destruction of the carriers of the 7.7 µm feature.

To further disentangle the complex interplay between PAH size, charge, structure, and the exciting radiation field, future observations are required. Obtaining MIRI MRS spectra covering the full 5–28 µm range for these and other targets would be invaluable. The inclusion of the 6.2 µm feature would provide an additional probe of the cation population, while the 17 µm feature would trace the largest, most stable PAHs, providing a more robust anchor for size distribution models. This study demonstrates the immense potential of JWST spectroscopy for dissecting the microphysics of the ISM and motivates such future investigations.

## 5. CONCLUSIONS

We have presented a detailed analysis of JWST/MIRI MRS mid-infrared spectra for four H II regions in the nearby grand-design spiral galaxy NGC 5457. Our investigation focused on the diagnostic power of the prominent PAH emission features at 7.7, 8.6, and 11.3 µm to probe the physical conditions and the processing of the interstellar PAH population in environments of active, high-mass star formation. The principal conclusions of this work are as follows:

Significant Environmental Diversity: The four H II regions exhibit a wide range of PAH spectral properties. Diagnostic flux ratios, such as F(8.6)/F(11.3), vary by up to a factor of three across the sample. This indicates that even within a single galaxy, H II regions are not a uniform class but encompass a broad spectrum of physical conditions.

Ionization as the Primary Driver: The observed variations in the PAH spectra are primarily driven by changes in the average ionization state of the PAH population. Using the F(8.6)/F(11.3) ratio as a proxy for ionization, we find that our sample spans from predominantly neutral, PDR-like environments to highly ionized regions directly exposed to intense FUV radiation.

A New Diagnostic for PAH Processing: We identify a strong anti-correlation between the F(7.7)/F(8.6) ratio and the F(8.6)/F(11.3) ionization tracer. This relationship reveals that the same energetic radiation that ionizes the PAHs also systematically alters the intrinsic emission properties of the surviving PAH cations.

An Evolutionary Sequence: The observed trend in our diagnostic diagram (F(7.7)/F(8.6) vs. F(8.6)/F(11.3)) is interpreted as an evolutionary or environmental sequence of PAH processing. As the radiation field hardens, PAHs become more ionized (increasing F(8.6)/F(11.3)), and the cation population is simultaneously altered, leading to a decrease in the F(7.7)/F(8.6) ratio. This provides a new, quantitative spectral signature of PAH processing and potential destruction in the harsh environments close to massive stars.

The unprecedented sensitivity and spatial resolution of JWST have opened a new window into the life cycle of PAHs. This work demonstrates how detailed, spatially resolved spectroscopy can untangle the complex interplay between these crucial molecules and their environment, providing a foundation for a more complete, observationally grounded model of the interstellar medium.

## COMPETING INTERESTS

The authors have declared that no competing interests exist.

## AUTHORS' CONTRIBUTIONS

AKS designed the study, performed the statistical analysis, and wrote the first draft of the manuscript. RKA managed the analyses of the study and managed the literature searches. SR read and approved the final manuscript.

## REFERENCES

Anand, R. K., Rastogi, S., & Kumar, B. (2023). PAH emission features in star-forming regions and late type stars. Journal of Astrophysics and Astronomy, 44(1), 47. https://doi.org/10.1007/s12036-023-09941-z

Anand, R. K., Singh, A. K., Sharma, S., Kumar, B., & Rastogi, S. (2025). Probing dust and PAH chemistry in evolved carbon-rich nebulae through optical and infrared observations. arXiv. https://doi.org/10.48550/arXiv.2508.04989

Argyriou, I., Glasse, A., Law, D. R., Labiano, Á., Álvarez-Márquez, J., Patapis, P., Kavanagh, P., Gasman, D., Mueller, M., Larson, K. L., Vandenbussche, B., Glauser, A. M., Royer, P., Dicken, D., Harkett, J., Sargent, B. A., Engesser, M., Jones, O., Kendrew, S., … Wells, M. (2023). JWST MIRI flight performance: The Medium-Resolution Spectrometer. *Astronomy and Astrophysics*, *675*. https://doi.org/10.1051/0004-6361/202346489

Bauschlicher, C. W., Peeters, E., & Allamandola, L. J. (2008). The Infrared Spectra of Very Large, Compact, Highly Symmetric, Polycyclic Aromatic Hydrocarbons (PAHs). *The Astrophysical Journal*, *678*(1), 316. https://doi.org/10.1086/533424

Belfiore, F., Leroy, A. K., Williams, T. G., Barnes, A., Bigiel, F., Boquien, M., Cao, Y., Chastenet, J., Congiu, E., Dale, D. A., Egorov, O. V., Eibensteiner, C., Emsellem, É., Glover, S. C. O., Groves, B., Hassani, H. R., Klessen, R. S., Kreckel, K., Neumann, L., … Watkins, E. J. (2023). Calibrating mid-infrared emission as a tracer of obscured star formation on HII-region scales in the era of JWST. *arXiv (Cornell University)*. https://doi.org/10.48550/arxiv.2306.11811

Berné, O., Foschino, S., Jalabert, F., & Joblin, C. (2022). Contribution of polycyclic aromatic hydrocarbon ionization to neutral gas heating in galaxies: model versus observations. *Astronomy and Astrophysics*, *667*. https://doi.org/10.1051/0004-6361/202243171

Bouwman, J., Kendrew, S., Greene, T. P., Bell, T. J., Lagage, P.-O., Schreiber, J., Dicken, D., Sloan, G. C., Espinoza, N., Scheithauer, S., Coulais, A., Fox, O. D., Gastaud, R., Glauser, A. M., Jones, O., Labiano, Á., Lahuis, F., Morrison, J., Murray, K. T., … Rieke, G. H. (2023). Spectroscopic Time Series Performance of the Mid-infrared Instrument on the JWST. *Publications of the Astronomical Society of the Pacific*, *135*(1045), 38002. https://doi.org/10.1088/1538-3873/acbc49

C., Jr. K. R., & Garnett, D. R. (1996). The Composition Gradient in M101 Revisited. I. H II Region Spectra and Excitation Properties. *The Astrophysical Journal*, *456*, 504. https://doi.org/10.1086/176675

Candian, A., & Sarre, P. J. (2015). The 11.2 $µ$m emission of PAHs in astrophysical objects. *arXiv*. https://doi.org/10.48550/ARXIV.1501.06811

Coy, B. P., Nixon, C. A., Rowe-Gurney, N., Achterberg, R. K., Lombardo, N. A., Fletcher, L. N., & Irwin, P. G. J. (2023). Spitzer IRS Observations of Titan as a Precursor to JWST MIRI Observations. *The Planetary Science Journal*, *4*(6), 114. https://doi.org/10.3847/psj/acd10f

Dicken, D., García-Marín, M., Shivaei, I., Guillard, P., Libralato, M., Glasse, A., Gordon, K. D., Cossou, C., Kavanagh, P., Temim, T., Flagey, N., Klaassen, P., Rieke, G. H., Wright, G., Alberts, S., Azzollini, R., Álvarez-Márquez, J.,

Bouchet, P., Bright, S. N., … Tamas, L. H. (2024). JWST MIRI flight performance: Imaging. *Astronomy and Astrophysics*, *689*. https://doi.org/10.1051/0004-6361/202449451

Dontot, L., Spiegelman, F., Zamith, S., & Rapacioli, M. (2020). Dependence upon charge of the vibrational spectra of small Polycyclic Aromatic Hydrocarbon clusters: the example of pyrene. *The European Physical Journal D*, *74*(11). https://doi.org/10.1140/epjd/e2020-10081-0

Draine, B. T., & Li, A. (2001). Infrared Emission from Interstellar Dust. I. Stochastic Heating of Small Grains. *The Astrophysical Journal*, *551*(2), 807. https://doi.org/10.1086/320227

Draine, B. T., Li, A., Hensley, B. S., Hunt, L. K., Sandström, K., & Smith, J. D. (2021). Excitation of Polycyclic Aromatic Hydrocarbon Emission: Dependence on Size Distribution, Ionization, and Starlight Spectrum and Intensity. *The Astrophysical Journal*, *917*(1), 3. https://doi.org/10.3847/1538-4357/abff51

Egorov, O. V., Kreckel, K., Sandström, K., Leroy, A. K., Glover, S. C. O., Groves, B., Kruijssen, J. M. D., Barnes, A., Belfiore, F., Bigiel, F., Blanc, G. A., Boquien, M., Cao, Y., Chastenet, J., Chevance, M., Congiu, E., Dale, D. A., Emsellem, É., Grasha, K., … Williams, T. G. (2023). PHANGS–JWST First Results: Destruction of the PAH Molecules in H ii Regions Probed by JWST and MUSE. *The Astrophysical Journal Letters*, *944*(2). https://doi.org/10.3847/2041-8213/acac92

Esposito, V. J., Ferrari, P., Buma, W. J., Fortenberry, R. C., Boersma, C., Candian, A., & Tielens, A. G. G. M. (2024). The infrared absorption spectrum of phenylacetylene and its deuterated isotopologue in the mid- to far-IR. *The Journal of Chemical Physics*, *160*(11). https://doi.org/10.1063/5.0191404

Esteban, C., Bresolin, F., García-Rojas, J., & Cipriano, L. T. S. (2019). Carbon, nitrogen and oxygen abundance gradients in M101 and M31. *Monthly Notices of the Royal Astronomical Society*. https://doi.org/10.1093/mnras/stz3134

Galliano, F. (2011). Dialectics of the PAH Abundance Trend with Metallicity. *EAS Publications Series*, *46*, 43. https://doi.org/10.1051/eas/1146004

Galliano, F., Madden, S. C., Tielens, A. G. G. M., Peeters, E., & Jones, A. P. (2008). Variations of the Mid-IR Aromatic Features inside and among Galaxies. *The Astrophysical Journal*, *679*(1), 310. https://doi.org/10.1086/587051

Galliano, F., Madden, S. C., Tielens, A. G. G. M., Peeters, E., & Jones, A. P. (2008). Variations of the Mid-IR Aromatic Features inside and among Galaxies. *The Astrophysical Journal*, *679*(1), 310. https://doi.org/10.1086/587051

García-Bernete, I., Rigopoulou, D., Alonso-Herrero, A., Donnan, F. R., Roche, P. F., Pereira-Santella, M., Labiano, Á., Arriba, L. P. de, Izumi, T., Almeida, C. R., Shimizu, T., Hönig, S., García-Burillo, S., Rosario, D. J., Ward, M. J., Bellocchi, E., Hicks, E. K. S., Fuller, L., & Packham, C. (2022). A high angular resolution view of the PAH emission in Seyfert galaxies using JWST/MRS data. *arXiv (Cornell University)*. https://doi.org/10.48550/arXiv.2208.11620

Goicoechea, J. R., Pety, J., Gérin, M., Hily-Blant, P., & Bourlot, J. L. (2009). The ionization fraction gradient across the Horsehead edge: an archetype for molecular clouds. *Astronomy and Astrophysics*, *498*(3), 771. https://doi.org/10.1051/0004-6361/200811496

Gordon, K. D., Engelbracht, C. W., Rieke, G. H., Misselt, K. A., Smith, J.-D. T., & Kennicutt, R. C. (2008). The Behavior of the Aromatic Features in M101 HII Regions: Evidence for Dust Processing. *arXiv*. https://doi.org/10.48550/ARXIV.0804.3223

Hélou, G., Malhotra, S., Hollenbach, D. J., Dale, D. A., & Contursi, A. (2000). Evidence for the Heating of Atomic Interstellar Gas by PAHs. *arXiv (Cornell University)*. https://doi.org/10.48550/arXiv.0010377

Hensley, B. S., Murray, C. E., & Dodici, M. (2022). Polycyclic Aromatic Hydrocarbons, Anomalous Microwave Emission, and their Connection to the Cold Neutral Medium. *The Astrophysical Journal*, *929*(1), 23. https://doi.org/10.3847/1538-4357/ac5cbd

Hu, N., Wang, E., Lin, Z., Kong, X., Cheng, F. Z., Zou, F., Fang, G., Lin, L., Mao, Y.-W., Wang, J., Zhou, X., Zhou, Z., Zhu, Y.-N., & Zou, H. (2018). M101: Spectral Observations of H ii Regions and Their Physical Properties. *The Astrophysical Journal*, *854*(1), 68. https://doi.org/10.3847/1538-4357/aaa6ca

Kang, X., Chang, R., Kudritzki, R., Gong, X.-B., & Zhang, F. (2021). Breaking the degeneracy between gas inflow and outflows with stellar metallicity: insights on M 101. *Monthly Notices of the Royal Astronomical Society*, *502*(2), 1967. https://doi.org/10.1093/mnras/stab147

Khramtsova, M. S., Wiebe, D. S., Boley, P. A., & Pavlyuchenkov, Ya. N. (2013b). Polycyclic aromatic hydrocarbons in spatially resolved extragalactic star-forming complexes. *Monthly Notices of the Royal Astronomical Society*, *431*(2), 2006. https://doi.org/10.1093/mnras/stt319

Khramtsova, M. S., Wiebe, D. S., Lozinskaya, T. A., & Egorov, O. V. (2014). Optical and infrared emission of H ii complexes as a clue to the PAH life cycle. *Monthly Notices of the Royal Astronomical Society*, *444*(1), 757. https://doi.org/10.1093/mnras/stu1482

Khramtsova, M. S., Wiebe, D. S., Lozinskaya, T. A., & Egorov, O. V. (2014). Optical and infrared emission of H ii complexes as a clue to the PAH life cycle. *Monthly Notices of the Royal Astronomical Society*, *444*(1), 757. https://doi.org/10.1093/mnras/stu1482

Knight, C., Peeters, E., Tielens, A. G. G. M., & Vacca, W. D. (2021). Characterizing the PAH emission in the Orion Bar. *Monthly Notices of the Royal Astronomical Society*, *509*(3), 3523. https://doi.org/10.1093/mnras/stab3047
Knight, C., Peeters, E., Wolfire, M. G., & Stock, D. J. (2021). Characterizing spatial variations of PAH emission in the reflection nebula NGC 1333. *Monthly Notices of the Royal Astronomical Society*, *510*(4), 4888. https://doi.org/10.1093/mnras/stab3295

Knight, C., Peeters, E., Wolfire, M. G., & Stock, D. J. (2021). Characterizing spatial variations of PAH emission in the reflection nebula NGC 1333. *Monthly Notices of the Royal Astronomical Society*, *510*(4), 4888. https://doi.org/10.1093/mnras/stab3295

Labiano, Á., Argyriou, I., Álvarez-Márquez, J., Glasse, A., Glauser, A. M., Patapis, P., Law, D. R., Brandl, B. R., Justtanont, K., Lahuis, F., Martínez-Galarza, J. R., Mueller, M., Noriega-Crespo, A., Royer, P., Shaughnessy, B., & Vandenbussche, B. (2021). Wavelength calibration and resolving power of the JWST MIRI Medium Resolution Spectrometer. *Astronomy and Astrophysics*, *656*. https://doi.org/10.1051/0004-6361/202140614

Li, A. (2020). Spitzer's perspective of polycyclic aromatic hydrocarbons in galaxies. *Nature Astronomy*, *4*(4), 339. https://doi.org/10.1038/s41550-020-1051-1

Lin, L., Zou, H., Kong, X., Lin, X., Mao, Y.-W., Cheng, F. Z., Jiang, Z., & Zhou, X. (2013). GRADIENTS OF STELLAR POPULATION PROPERTIES AND EVOLUTION CLUES IN A NEARBY GALAXY M101. *The Astrophysical Journal*, *769*(2), 127. https://doi.org/10.1088/0004-637x/769/2/127

Lin, L., Zou, H., Kong, X., Lin, X., Mao, Y., Jiang, Z., & Zhou, X. (2013). Gradients of Stellar Population Properties and Evolution Clues in a Nearby Galaxy M 101. *arXiv*. https://doi.org/10.48550/ARXIV.1304.3017

Linden, S. T., Pérez, G., Calzetti, D., Maji, S., Messa, M., Whitmore, B., Chandar, R., Adamo, A., Grasha, K., Cook, D., Elmegreen, B. G., Dale, D. A., Sacchi, E., Sabbi, E., Grebel, E. K., & Smith, L. J. (2022). Star Cluster Formation and Evolution in M101: An Investigation with the Legacy Extragalactic UV Survey. *The Astrophysical Journal*, *935*(2), 166. https://doi.org/10.3847/1538-4357/ac7c07

Mackie, C. J., Peeters, E., Bauschlicher, C. W., & Cami, J. (2015). CHARACTERIZING THE INFRARED SPECTRA OF SMALL, NEUTRAL, FULLY DEHYDROGENATED POLYCYCLIC AROMATIC HYDROCARBONS. *The Astrophysical Journal*, *799*(2), 131. https://doi.org/10.1088/0004-637x/799/2/131

Maltseva, E., Petrignani, A., Candian, A., Mackie, C. J., Huang, X., Lee, T. J., Tielens, A. G. G. M., Oomens, J., & Buma, W. J. (2016). HIGH-RESOLUTION IR ABSORPTION SPECTROSCOPY OF POLYCYCLIC AROMATIC HYDROCARBONS IN THE 3 μm REGION: ROLE OF PERIPHERY. *The Astrophysical Journal*, *831*(1), 58. https://doi.org/10.3847/0004-637x/831/1/58

Maurya, A., Singh, R., & Rastogi, S. (2023). Study of vibrational spectra of polycyclic aromatic hydrocarbons with phenyl side group. Journal of Molecular Spectroscopy, 391, 111720. https://doi.org/10.1016/j.jms.2022.111720

Maurya, A., & Rastogi, S. (2015). Vibrational spectroscopic study of vinyl substituted polycyclic aromatic hydrocarbons. Spectrochimica Acta Part A: Molecular and Biomolecular Spectroscopy, 151, 1–10. https://doi.org/10.1016/j.saa.2015.06.069

Maragkoudakis, A., Boersma, C., Temi, P., Bregman, J. D., & Allamandola, L. J. (2022). Linking Characteristics of the Polycyclic Aromatic Hydrocarbon Population with Galaxy Properties: A Quantitative Approach Using the NASA Ames PAH IR Spectroscopic Database. *The Astrophysical Journal*, *931*(1), 38. https://doi.org/10.3847/1538-4357/ac666f

Maragkoudakis, A., Peeters, E., & Ricca, A. (2020). Probing the size and charge of polycyclic aromatic hydrocarbons. *Monthly Notices of the Royal Astronomical Society*, *494*(1), 642. https://doi.org/10.1093/mnras/staa681

Maragkoudakis, A., Peeters, E., & Ricca, A. (2023). Spectral variations among different scenarios of PAH processing or formation. *Monthly Notices of the Royal Astronomical Society*, *520*(4), 5354. https://doi.org/10.1093/mnras/stad465

Maragkoudakis, A., Peeters, E., Ricca, A., & Boersma, C. (2023). Polycyclic aromatic hydrocarbon size tracers. *Monthly Notices of the Royal Astronomical Society*, *524*(3), 3429. https://doi.org/10.1093/mnras/stad2062

Micelotta, E. R., Jones, A. P., & Tielens, A. G. G. M. (2009). Polycyclic aromatic hydrocarbon processing in a hot gas. *Astronomy and Astrophysics*, *510*. https://doi.org/10.1051/0004-6361/200911683

Micelotta, E. R., Jones, A. P., & Tielens, A. G. G. M. (2010). Polycyclic aromatic hydrocarbon processing by cosmic rays. *Astronomy and Astrophysics*, *526*. https://doi.org/10.1051/0004-6361/201015741

Monfredini, T., Quitián-Lara, H. M., Fantuzzi, F., Wolff, W., Mendoza, E., Lago, A. F., Sales, D. A., Pastoriza, M. G., & Boechat-Roberty, H. M. (2019). Destruction and multiple ionization of PAHs by X-rays in circumnuclear regions of AGNs. *Monthly Notices of the Royal Astronomical Society*, *488*(1), 451. https://doi.org/10.1093/mnras/stz1021

Morrison, J., Dicken, D., Argyriou, I., Ressler, M. E., Gordon, K. D., Regan, M. W., Cracraft, M., Rieke, G. H., Engesser, M., Alberts, S., Álvarez-Márquez, J., Colbert, J., Fox, O. D., Gasman, D., Law, D. R., García-Marín, M., Gáspár, A., Guillard, P., Kendrew, S., … Sloan, G. C. (2023). JWST MIRI Flight Performance: Detector Effects and Data Reduction Algorithms. *Publications of the Astronomical Society of the Pacific*, *135*(1049), 75004. https://doi.org/10.1088/1538-3873/acdea6

Murata, K. L., Yamada, R., Oyabu, S., Kaneda, H., Ishihara, D., Yamagishi, M., Kokusho, T., & Takeuchi, T. T. (2017). A Relationship of Polycyclic Aromatic Hydrocarbon Features with Galaxy Merger in Star-forming Galaxies at $z<0.2$. *arXiv*. https://doi.org/10.48550/ARXIV.1707.09358

Pathak, A., & Rastogi, S. (2006). Computational study of neutral and cationic pericondensed polycyclic aromatic hydrocarbons. *Chemical Physics, 326*(2–3), 315–328. https://doi.org/10.1016/j.chemphys.2006.02.008

Pathak, A., & Rastogi, S. (2008). Modeling the interstellar aromatic infrared bands with co-added spectra of PAHs. Astronomy & Astrophysics, 485(3), 735–742. https://doi.org/10.1051/0004-6361:20066618

Patapis, P., Argyriou, I., Law, D. R., Glauser, A. M., Glasse, A., Labiano, Á., Álvarez-Márquez, J., Kavanagh, P., Gasman, D., Mueller, M., Larson, K. L., Vandenbussche, B., Lee, D., Klaassen, P., Guillard, P., & Wright, G. (2023). Geometric distortion and astrometric calibration of the JWST MIRI Medium Resolution Spectrometer. *Astronomy and Astrophysics*, *682*. https://doi.org/10.1051/0004-6361/202347339

Pavlyuchenkov, Ya. N., Kirsanova, M. S., & Wiebe, D. S. (2013). Infrared emission and the destruction of dust in HII regions. *Astronomy Reports*, *57*(8), 573. https://doi.org/10.1134/s1063772913070056

Pavlyuchenkov, Ya. N., Wiebe, D. S., Akimkin, V., Khramtsova, M. S., & Henning, Th. (2012). Stochastic grain heating and mid-infrared emission in protostellar cores. *Monthly Notices of the Royal Astronomical Society*, *421*(3), 2430. https://doi.org/10.1111/j.1365-2966.2012.20480.x

Peeters, E., Hony, S., Van Kerckhoven, C., Tielens, A. G. G. M., Allamandola, L. J., Hudgins, D. M., & Bauschlicher, C. W., Jr. (2002). The rich 6 to 9 μm spectrum of interstellar PAHs. Astronomy & Astrophysics, 390(3), 1089–1113. https://doi.org/10.1051/0004-6361:20020773

Peeters, E., Spoon, H. W. W., & Tielens, A. G. G. M. (2004). PAHs as a tracer of star formation. *arXiv (Cornell University)*. http://arxiv.org/abs/astro-ph/0406183

Pereira-Santaella, M., Álvarez-Márquez, J., García-Bernete, I., Labiano, Á., Colina, L., Alonso-Herrero, A., Bellocchi, E., García-Burillo, S., Hönig, S. F., Almeida, C. R., & Rosario, D. J. (2022). Low-power jet-ISM interaction in NGC 7319 revealed by JWST/MIRI MRS. *arXiv (Cornell University)*. https://doi.org/10.48550/arxiv.2208.04835

Pleuss, P. O., Heller, C., & Fricke, K. J. (2000). The Impact of Resolution on Observed HII Region Properties from WFPC2 Observations of M101. *arXiv (Cornell University)*. https://doi.org/10.48550/arXiv.0010565

Rau, S.-J., Hirashita, H., & Murga, M. S. (2019). Modelling the evolution of PAH abundance in galaxies. *Monthly Notices of the Royal Astronomical Society*, *489*(4), 5218. https://doi.org/10.1093/mnras/stz2532

Rigopoulou, D., Donnan, F. R., García-Bernete, I., Pereira-Santaella, M., Alonso-Herrero, A., Davies, R., Hunt, L. K., Roche, P. F., & Shimizu, T. (2024a). Polycyclic Aromatic Hydrocarbon Emission in Galaxies as seen with JWST. *arXiv (Cornell University)*. https://doi.org/10.48550/arxiv.2406.11415

Rigopoulou, D., Donnan, F. R., García-Bernete, I., Pereira-Santaella, M., Alonso-Herrero, A., Davies, R., Hunt, L. K., Roche, P. F., & Shimizu, T. (2024b). Polycyclic aromatic hydrocarbon emission in galaxies as seen with JWST. *Monthly Notices of the Royal Astronomical Society*, *532*(2), 1598. https://doi.org/10.1093/mnras/stae1535

Rigopoulou, D., Donnan, F. R., García-Bernete, I., Pereira-Santaella, M., Alonso-Herrero, A., Davies, R., Hunt, L. K., Roche, P. F., & Shimizu, T. (2024). Polycyclic aromatic hydrocarbon emission in galaxies as seen with JWST. *Monthly Notices of the Royal Astronomical Society*, *532*(2), 1598. https://doi.org/10.1093/mnras/stae1535

Rogers, N. S. J., Arellano-Córdova, K. Z., Berg, D., Pogge, R. W., & Skillman, E. D. (2023, May). The novel MIR abundance diagnostic Ne23 (JWST Proposal. Cycle 2, ID. #4297). https://ui.adsabs.harvard.edu/abs/2023jwst.prop.4297R

Sakon, I., Onaka, T., OKAMOTO, Y. K., Kataza, H., Kaneda, H., & Honda, M. (2011). IONIZATION OF POLYCYCLIC AROMATIC HYDROCARBON MOLECULES AROUND THE HERBIG Ae/Be ENVIRONMENT. In *World Scientific Publishing Co. Pte. Ltd. eBooks* (p. 143). https://doi.org/10.1142/9789812708922_0014

Sandstrom, K. M., Bolatto, A. D., Draine, B., Bot, C., & Stanimirovic, S. (2010). The Spitzer Survey of the Small Magellanic Cloud (S3MC): Insights into the Life-Cycle of Polycyclic Aromatic Hydrocarbons. *arXiv*. https://doi.org/10.48550/ARXIV.1003.4516

Sandström, K., Koch, E. W., Leroy, A. K., Rosolowsky, E., Emsellem, É., Smith, R. J., Egorov, O. V., Williams, T. G., Larson, K. L., Lee, J., Schinnerer, E., Thilker, D. A., Barnes, A., Belfiore, F., Bigiel, F., Blanc, G. A., Bolatto, A. D., Boquien, M., Cao, Y., … Watkins, E. J. (2023). PHANGS–JWST First Results: Tracing the Diffuse Interstellar Medium with JWST Imaging of Polycyclic Aromatic Hydrocarbon Emission in Nearby Galaxies. *The Astrophysical Journal Letters*, *944*(2). https://doi.org/10.3847/2041-8213/aca972

Semmler, J., Yang, P., & Crawford, G. (1991). Gas chromatography/Fourier transform infrared studies of gas-phase polynuclear aromatic hydrocarbons. *Vibrational Spectroscopy*, *2*(4), 189. https://doi.org/10.1016/0924-2031(91)85026-j

Sidhu, A., Tielens, A. G. G. M., Peeters, E., & Cami, J. (2022). Polycyclic Aromatic Hydrocarbon emission model in photodissociation regions I: Application to the 3.3, 6.2, and 11.2 $\mu$m bands. *arXiv*. https://doi.org/10.48550/ARXIV.2205.03304

Sidhu, A., Tielens, A. G. G. M., Peeters, E., & Cami, J. (2022). Polycyclic Aromatic Hydrocarbon emission model in photodissociation regions I: Application to the 3.3, 6.2, and 11.2 $\mu$m bands. *arXiv*. https://doi.org/10.48550/ARXIV.2205.03304

Siebenmorgen, R. (2009). PAH destruction and survival in the disks of T Tauri stars. *arXiv (Cornell University)*. https://doi.org/10.48550/arxiv.0911.4360

Sutter, J., Sandstrom, K., Chastenet, J., Leroy, A. K., Koch, E. W., Williams, T. G., Chown, R., Belfiore, F., Bigiel, F., Boquien, M., Cao, Y., Chevance, M., Dale, D. A., Egorov, O. V., Glover, S. C. O., Groves, B., Klessen, R. S., Kreckel, K., Larson, K. L., … Watkins, E. J. (2024). *The Fraction of Dust Mass in the Form of PAHs on 10-50 pc Scales in Nearby Galaxies*. https://doi.org/10.48550/ARXIV.2405.15102

Tielens, A. G. G. M. (2008). Interstellar polycyclic aromatic hydrocarbon molecules. Annual Review of Astronomy and Astrophysics, 46(1), 289–337. https://doi.org/10.1146/annurev.astro.46.060407.145211

Ujjwal, K., Kartha, S. S., Akhil, K. R., Mathew, B., Subramanian, S., Sudheesh, T. P., & Thomas, R. (2024). Disentangling the association of PAH molecules with star formation. *Astronomy and Astrophysics*, *684*. https://doi.org/10.1051/0004-6361/202347620

Wenzel, G., Jiménez-Redondo, M., Ončák, M., McGuire, B. A., Brünken, S., Caselli, P., & Jusko, P. (2025). Infrared Spectroscopy of Pentagon-Containing PAHs: Indenyl and Fluorenyl Anions and Indenyl Cation. *The Journal of Physical Chemistry Letters*, 3938. https://doi.org/10.1021/acs.jpclett.5c00570

Whitcomb, C. M., Sandström, K., Leroy, A. K., & Smith, J. D. (2023). Star Formation and Molecular Gas Diagnostics with Mid- and Far-infrared Emission. *The Astrophysical Journal*, *948*(2), 88. https://doi.org/10.3847/1538-4357/acc316

Whitcomb, C. M., Smith, J. D., Sandström, K., Starkey, C. A., Donnelly, G. P., Draine, B. T., Skillman, E. D., Dale, D. A., Armus, L., Hensley, B. S., Lai, T. S.-Y., & Kennicutt, R. C. (2024). The Metallicity Dependence of PAH Emission in Galaxies. I. Insights from Deep Radial Spitzer Spectroscopy. *The Astrophysical Journal*, *974*(1), 20. https://doi.org/10.3847/1538-4357/ad66c8

Wright, G., Rieke, G. H., Glasse, A., Ressler, M. E., García-Marín, M., Aguilar, J., Alberts, S., Álvarez-Márquez, J., Argyriou, I., Banks, K. C., Baudoz, P., Boccaletti, A., Bouchet, P., Bouwman, J., Brandl, B. R., Breda, D., Bright, S. N., Cale, S., Colina, L., … Weilert, M. (2023). The Mid-infrared Instrument for JWST and Its In-flight Performance. *Publications of the Astronomical Society of the Pacific*, *135*(1046), 48003. https://doi.org/10.1088/1538-3873/acbe66

Zang, R. X., Maragkoudakis, A., & Peeters, E. (2022). The spatially resolved PAH characteristics in the Whirlpool Galaxy (M51a). *Monthly Notices of the Royal Astronomical Society*, *511*(4), 5142. https://doi.org/10.1093/mnras/stac214

Zhen, J., Castillo, S. R., Joblin, C., Mulas, G., Sabbah, H., Giuliani, A., Nahon, L., Martin, S., Champeaux, J. P., & Mayer, P. (2016). VUV PHOTO-PROCESSING OF PAH CATIONS: QUANTITATIVE STUDY ON THE IONIZATION VERSUS FRAGMENTATION PROCESSES. *The Astrophysical Journal*, *822*(2), 113. https://doi.org/10.3847/0004-637x/822/2/113